# Quantitative mapping of smooth topographic landscapes produced by thermal scanning-probe lithography

Camilla H. Sørensen*, Magnus V. Nielsen*, Sander J. Linde*, Duc Hieu Nguyen, Christoffer E. Iversen, Robert Jensen, Søren Raza, Peter Bøggild, Timothy J. Booth, Nolan Lassaline

2DPhys Laboratory, Department of Physics, Technical University of Denmark, 2800 Kongens Lyngby, Denmark

*These authors contributed equally

**Scanning probe microscopy (SPM) is a powerful technique for mapping nanoscale surface properties through tip–sample interactions. Thermal scanning-probe lithography (tSPL) is an advanced SPM variant that uses a silicon tip on a heated cantilever to sculpt and measure polymer films with nanometer precision. The surfaces produced by tSPL—smooth topographic landscapes—allow mathematically defined contours to be fabricated on the nanoscale, enabling sophisticated functionalities for photonic, electronic, chemical, and biological technologies. Evaluating the physical effects of a landscape requires fitting arbitrary mathematical functions to SPM datasets, however, this capability does not exist in standard analysis programs. Here, we provide an open-source software package (*FunFit*) to fit analytical functions to SPM data and develop a fabrication and characterization protocol based on this analysis. We demonstrate the benefit of this approach by patterning periodic and quasiperiodic landscapes in a polymer resist with tSPL, which we transfer to hexagonal boron nitride (hBN) flakes with high fidelity via reactive-ion etching. The topographic landscapes in polymers and hBN are measured with tSPL and atomic force microscopy (AFM), respectively. Within the *FunFit* program, the datasets are corrected for artefacts, fit with analytical functions, and compared, providing critical feedback on the fabrication procedure. Beyond application to tSPL, this protocol can improve analysis, reproducibility, and process development for a broad range of SPM experiments. The protocol can be performed within a working day by an inexperienced user, where fabrication and characterization take a few hours and software analysis takes a few minutes.**

# Introduction

Scanning probe microscopy (SPM) has revolutionized science by allowing nanoscale surface imaging with high (or even atomic) resolution[1]. Beyond imaging, scanning probes can be used as lithographic instruments to structure and manipulate materials, offering a versatile toolbox for patterning and characterization on the nanoscale[2,3]. Thermal scanning-probe lithography (tSPL)[4] is an advanced SPM technique that uses a silicon tip on a heated cantilever to locally sublimate polymer films[5,6]. The key advantage of tSPL is the ability to control the temperature and applied force of the tip on the sample surface, which can be used to sculpt smooth topographic landscapes with nanometer precision[7,8]. Recently, these landscapes have been leveraged to create smooth, mathematically precise potentials to control the motion and behavior of photons[9], electrons[10,11], and nanoparticles[12], or producing structured tissue environments for biological studies[13], opening a new field of exploration for experimental nanoscience.

One of the primary challenges in tSPL research is to accurately and reliably transfer patterns from the polymer layer to a variety of other materials such as metals, semiconductors, and dielectrics, which is required for most applications[6]. Achieving high-quality pattern transfer requires a stable and reproducible process based on techniques such as deposition[11], etching[14], or templating[9]. Furthermore, the structured surfaces need to be analysed to produce a quantitative evaluation of the pattern fidelity both before and after the transfer process. There is growing interest in developing fabrication and characterization routines based on tSPL, however, standardized methodology is still lacking in this maturing field.

This protocol establishes simple guidelines[15] for patterning polymer surfaces with tSPL and transferring these patterns into nanomaterials. We demonstrate this by patterning polyphthalaldehyde (PPA) resist with tSPL and transferring these patterns with reactive-ion etching (RIE) to hexagonal boron nitride (hBN)[16], a layered crystalline material with technologically relevant properties for 2D materials and van der Waals (vdW) heterostructure engineering. While the protocol is developed for these materials specifically, the approach presented here can be easily adapted to other material systems[17] and fabrication processes[9]. To complement the fabrication procedure presented in the protocol, we also provide an open-source software package, *FunFit*, that treats SPM data by removing scan artefacts and fitting mathematical functions to measured surface profiles. This capability, which does not exist in standard analysis programs, allows topographic landscapes produced by tSPL to be quantified and properties such as roughness, patterning errors, and fidelity to be evaluated. Thus, *FunFit* equips the user with a tool to quantitatively evaluate and compare patterns in different stages of the process to facilitate reproducibility, accuracy, and optimization of the fabrication routine. Beyond tSPL landscapes, this protocol offers value as an advanced analysis procedure for a broad range of scanning probe experiments and fabrication processes where function fitting and detailed mathematical evaluation would be beneficial[18].

**Development of the protocol**

This protocol was developed to establish an experimental framework for tSPL patterning, pattern transfer, and pattern evaluation. The protocol is designed for non-experts, such that researchers entering the field can use the protocol as a guide when they begin tSPL research. The software package is user friendly and designed to improve pattern transfer, reproducibility, and process optimization for tSPL patterning[19] and other probe-based nanofabrication routines[2].

**Overview of the procedure**

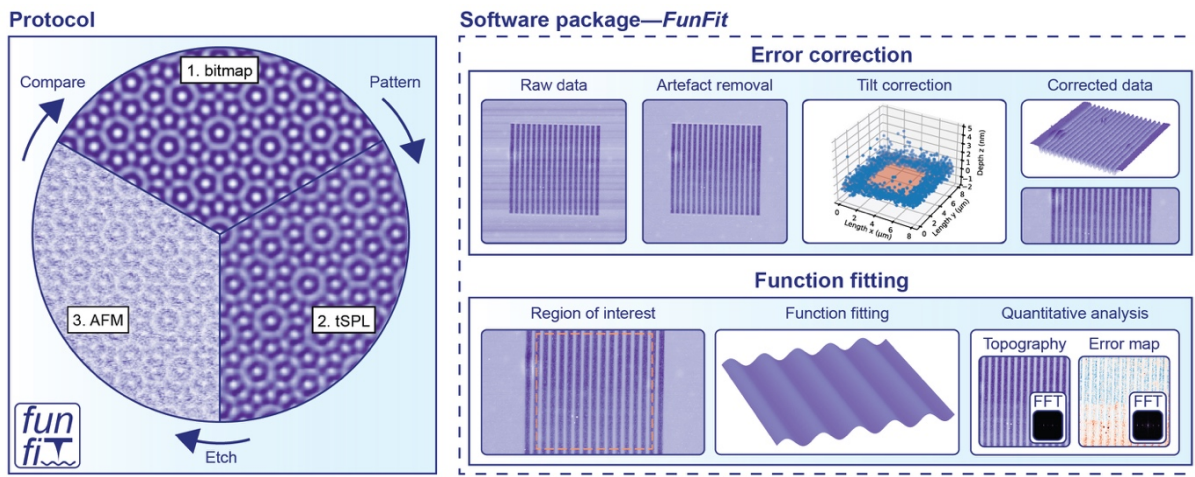

**Fig. 1 | Protocol overview. Left**: The protocol establishes guidelines for designing, patterning, and transferring smooth topographic landscapes produced by tSPL. We demonstrate this approach by patterning PPA resist and transferring the patterns to hBN using RIE. **Right**: The software package *FunFit* performs two main functions, error correction and function fitting, allowing quantitative analysis of tSPL patterns.

Protocol

The protocol provides step-by-step instructions on how to use tSPL to produce greyscale topographic landscapes in PPA based on mathematical design functions[15], and successive pattern transfer to hBN by RIE[10,20]. An overview of the protocol is illustrated in Fig. 1. In brief, the procedure contains the following three parts: 1) The target design function is used to generate a greyscale bitmap image (see Box 1). 2) tSPL is then used to replicate the surface in a PPA film, where the resulting topography is measured during patterning. 3) RIE is used to transfer the pattern to an underlying hBN flake, and the final topography is measured by AFM. Experimental details are found in the Experimental Design and Procedure sections.

<u>Software Package</u>

To assess and compare the generated patterns, the software package *FunFit* combines error correction with function fitting (See Fig. 1). The software package is divided into four main stages: 1) import of topography data (tSPL/AFM data), 2) data preprocessing to correct for scan artefacts and tilt, 3) function fitting, and 4) quantitative analysis based on the fitting results. This allows the user to quantify e.g. surface roughness and pattern characteristics, providing important information for process optimization and reproducibility. Details are found in the Procedure section.

**Applications of the method**

The protocol was specifically developed for patterning smooth topographic landscapes in PPA with tSPL and transferring these patterns to hBN with RIE. However, the protocol is more broadly useful for any scanning probe experiments and fabrication routines where function fitting and detailed mathematical evaluation is required, in both academic and industrial research environments.

**Alternative methods**

There are a variety of options for analyzing SPM data with software packages. Commercial programs, such as *NanoScope Analysis* from Bruker Corporation, offer user-friendly software that can perform basic SPM analysis with a broad library of functionalities. Open-source options, such as *Gwyddion*, offer an alternative platform for analyzing and visualizing SPM data with a number of user-inspired features. While the available analysis programs provide user-friendly packages and a broad variety of analysis features, they do not provide a workspace to fit arbitrary mathematical functions, as defined by the user. Our software protocol bridges this gap by combining data pre-processing and visualization with quantitative analysis based on fitting 2D analytical functions to topographic data.

**Advantages and limitations**

Key advantages:

- Combines data-preprocessing, visualization, and quantitative analysis based on function fitting. This is useful for characterization, reproducibility, process development and optimization of probe-based lab procedures.
- User friendly software offering easy and accessible SPM data analysis for beginners.

Key limitations:

- Requires SPM datasets with sufficient sampling and appropriate scan settings, where both patterned and flat reference regions are measured.
- Roughness from an etching process can make function fitting difficult, especially for intricate mathematical functions.

**Experimental design**

We demonstrate a valuable use case of the protocol by patterning intricate periodic and quasiperiodic landscapes in PPA with tSPL, which we transfer to hBN flakes with high fidelity using RIE. With minor modifications, the protocol can be easily extended to other tSPL patterns and alternative materials, such as semiconductors, metals, or other vdW crystals and heterostructures.

An overview of the sample fabrication is illustrated in Fig. 2 and details of the fabrication steps are provided in the Procedure section. In brief, a PPA polymer film is spin coated over a cleaned Si/SiO$_2$ chip on which exfoliated hBN crystal flakes reside. tSPL is used to pattern (and simultaneously measure) a mathematical function in the polymer surface, which is then transferred to the hBN surface with RIE. The remaining PPA layer is removed, and a final oxygen plasma cleaning is performed.

**Expertise needed to implement the protocol**

The user should be familiar with the basic operating principles of tSPL and simple preparation steps such as spin coating and optical microscopy. This protocol aims to provide the information and troubleshooting assistance necessary to obtain high quality patterns and pattern transfer, even for an inexperienced tSPL user. Moreover, it is assumed that the user is familiar with basic AFM characterization, general fume hood safety, and pattern-transfer processes such as dry etching (RIE)[14].

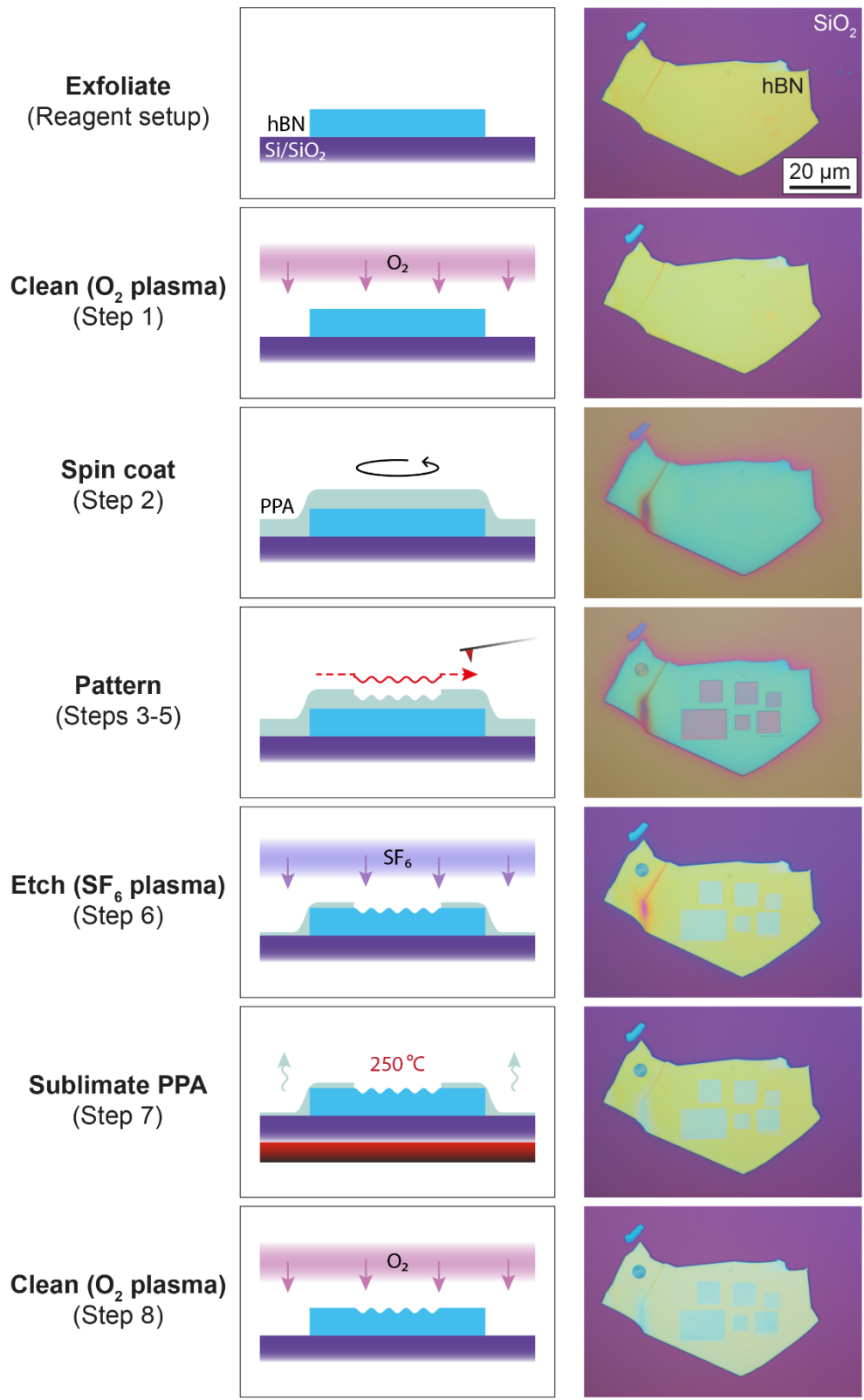

**Fig. 2 | Fabrication procedure. Left**: Process steps. The step numbers refer to the numbered steps in the Procedure section, where more detailed information is available. **Middle**: Illustration of process steps. **Right**: Optical microscope image of hBN flake on a SiO$_2$ surface at every process step.

## Materials

### Reagents

- Polyphthalaldehyde (PPA) powder (Pheonix 81, Allresist GmbH, 950k, linear).

▲ **CRITICAL** The PPA powder must be stored in the freezer (−18 °C)

▲ **CRITICAL** The PPA solution must be stored in the fridge (4 °C)

- Anisole (anhydrous 99.7%, 1L, Merck (Sigma-Aldrich), product no. 296295, https://www.sigmaaldrich.com/DK/en/product/sial/296295).

- Isopropanol (IPA, 2-propanol ≥99.8%, 2.5L, VWR/Avantor, cat. no. 20880.320, https://dk.vwr.com/store/product?keyword=20880.320).

- Adhesive tape (blue tape) (1007R Silicone-Free Blue Adhesive Plastic Film - Medium tack, 5.9", Semiconductor Production Systems (SPS), https://www.sps-international.com/product/1007r-silicone-free-blue-adhesive-plastic-film---medium-tack/7634/).

- Bulk hBN crystal (HQ Graphene, https://www.hqgraphene.com/h-BN.php).

- Si/SiO$_2$ (300 nm) wafer (Siegert Wafer, Part-No. Z14323 or Z14390. The latter allows the Si layer to be used as a back-gate for electrical devices. If this is irrelevant, the former suffices).

### Equipment

- NanoFrazor Scholar (Heidelberg Instruments Nano).
- MiniLab 026 Soft Etching system, RIE chamber (Moorfield Nanotechnology).
- EMS 6000 spin-coater (EMS).
- Dimension Icon-PT AFM (Bruker Corporation).
- Nikon Eclipse L200N optical microscope (Nikon).

- Vortex mixer (Heathrow Scientific Vortexer (multi-purpose), Fisher Scientific, cat. no. 15599811) or similar.
- Glass vials (Screw neck vial ND13, 4 mL, amber glass, VWR/Avantor, cat. no. 548-0052A, https://dk.vwr.com/store/product?keyword=548-0052A and PTFE cap.
- PTFE filters, pore size ≤ 0.2 µm (Corning, part no. CLS431227)
- Spatula.
- Cotton swab.
- Diamond-tipped pen.
- Hot plate.
- Balance.
- $N_2$ gun.

**Reagent setup: PPA solution**

PPA is purchased from Allresist GmbH and shipped in powder form in an air-tight container. The PPA powder should be stored in a freezer at −18 °C and removed only when making a new vial of PPA solution in anisole.

■ **PAUSE POINT** The solution can be stored for up to 1 year in a freezer at −18 °C.

▲ **CRITICAL STEP** The PPA powder must be removed from the freezer 1 hour before making the solution to bring the storage vial to room temperature and prevent moisture condensation.

The PPA powder should be transferred to a vial on a balance to measure the mass of PPA added. The amount of PPA required depends on the target concentration, which ultimately determines the thickness of a spin-coated film for a given spin speed.

▲ **CRITICAL STEP** The PPA powder should be transferred to a glass vial with a polytetrafluoroethylene (PTFE) cap, and the transfer tool should also be made of PTFE or glass to prevent contamination.

Here an example calculation is performed to make a 1 mL solution of PPA in anisole with a concentration of five weight per-cent (5 wt. %). A 5 wt. % film is expected to spin to a thickness range of ~40–100 nm for spin-coating speeds in the range of 2000–6000 r.p.m. The concentration in wt. % is given by:

$$\text{Concentration(wt. \%)} = \frac{m_{\text{PPA}}}{m_{\text{PPA}} + m_{\text{anisole}}} = 5\%,$$

where $m_{\text{PPA}}$ is the mass of PPA powder added to be the vial, and $m_{\text{anisole}}$ is the mass of anisole that will be added to prepare the solution. For a 1 mL solution of anisole with density $\rho = 0.995\,\frac{\text{g}}{\text{mL}}$,

$$m_{\text{anisole}} = \rho V = \left(0.995\,\frac{\text{g}}{\text{mL}}\right)(1\text{ mL}) = 0.995\text{ g}.$$

The mass of PPA to be added is then

$$m_{\text{PPA}} = \frac{m_{\text{anisole}} \times \text{Concentration(wt. \%)}}{1 - \text{Concentration(wt. \%)}} = \frac{(0.995\text{ g})(0.05)}{1 - 0.05} = 52\text{ mg}.$$

A spatula should be used to transfer 52 mg of PPA from its storage container to a glass vial on a balance. Once this is complete, the glass vial should be brought to the fume hood where 1 mL of anisole is added to the vial with a pipette. The vial should then be lightly shaken for 15 minutes ideally using a mechanical vortex mixer, but hand shaking will suffice as an alternative. At the end of the shaking process, the PPA powder should be fully dissolved, leaving no visible particles when inspected by the eye. The final step to ensure a clean PPA solution is to filter the contents of the vial through a porous PTFE filter with pore size $\leq$ 0.2 $\mu$m into another clean glass vial. This vial should then be left to rest another 1 hour before spin coating to allow air bubbles to settle.

■ **PAUSE POINT** The solution can be stored for up to 3 months in a fridge at 4 °C.

◆ **TROUBLESHOOTING** The PPA solution may last longer than 3 months, and degradation is difficult to detect as there are no obvious visual cues during the process steps. Thickness variations in a spin-coated film at a constant speed may indicate degradation. A degraded PPA solution will pose problems during tSPL patterning, characterized by a substantial difficulty in performing simple calibration patterns, and rapid/unpredictable changes in the adhesion length and patterning properties of the tip. These difficulties can, however, arise for different reasons with good PPA solutions as well, so it is recommended to use PPA within the first 3 months of preparation.

▲ **CRITICAL STEP** The PPA solution should be removed from the fridge 15 minutes before spin coating to bring the storage vial to room temperature and prevent moisture condensation inside the vial.

**Reagent setup: Mechanical exfoliation of hBN**

Mechanical exfoliation is used to produce high quality hBN flakes from bulk crystals on Si/SiO$_2$ chips.

Preparation and cleaning:

- Use a pen with a sharp diamond tip to cleave a Si/SiO$_2$ wafer into ~1.5 cm x 1.5 cm square chips.
- Place the Si/SiO$_2$ chip(s) in warm acetone in a glass beaker on a hotplate set to 60 °C for 5 min. To prevent evaporation during the cleaning process, cover the beaker opening with aluminium foil.

▲ **CAUTION** Warm acetone is above the flash point. This step should be performed in a fumehoood while wearing personal protective equipment. The hot plate should be certified to a standardized safety rating, such as ATEX in Europe.

- Rinse the chip(s) in isopropanol (IPA) immediately after.
- Blow-dry the chip(s) using a N$_2$ gun.

Fig. 3 presents a workflow for the mechanical exfoliation process, and detailed instructions for each subfigure are provided below:

a) Fold the edges of a long piece of blue tape, as this makes it easier to handle. Use a set of tweezers to place a small piece of bulk hBN crystal on the sticky side of the tape, tap lightly, and peel off to leave a thin crystal flake on the tape.

Fold the tape onto itself to make additional replicas of the original flake. It can be an advantage to do this in a structured fashion such that the copied flakes form ordered arrays. Later, this will help to keep track of the positions of specific exfoliated flakes if the Si/SiO$_2$ chips do not have markings. Cut the piece of tape in half to produce two stamp pieces.

b) Rub the stamps against new pieces of tape with fingers (1), and peel apart quickly (2), to produce additional stamps. This process gradually thins down the flakes. Repeat until all flakes appear equally thin on all stamps.

c) To transfer the crystal flakes from a stamp onto a Si/SiO$_2$ chip, place the stamp over the chip and use a cotton swab to gently rub the backside of the tape. Slowly peel off the tape (use ~2 min per chip) at a low angle ($\theta < 90°$).

d) The Si/ SiO$_2$ chip should now have exfoliated hBN crystal flakes on top.

■ **PAUSE POINT** The exfoliated hBN flakes can be stored for at least several months in a controlled atmosphere such as a glove box. In ambient conditions (air), a self-organizing organic contamination layer has been reported to form after a few days[21].

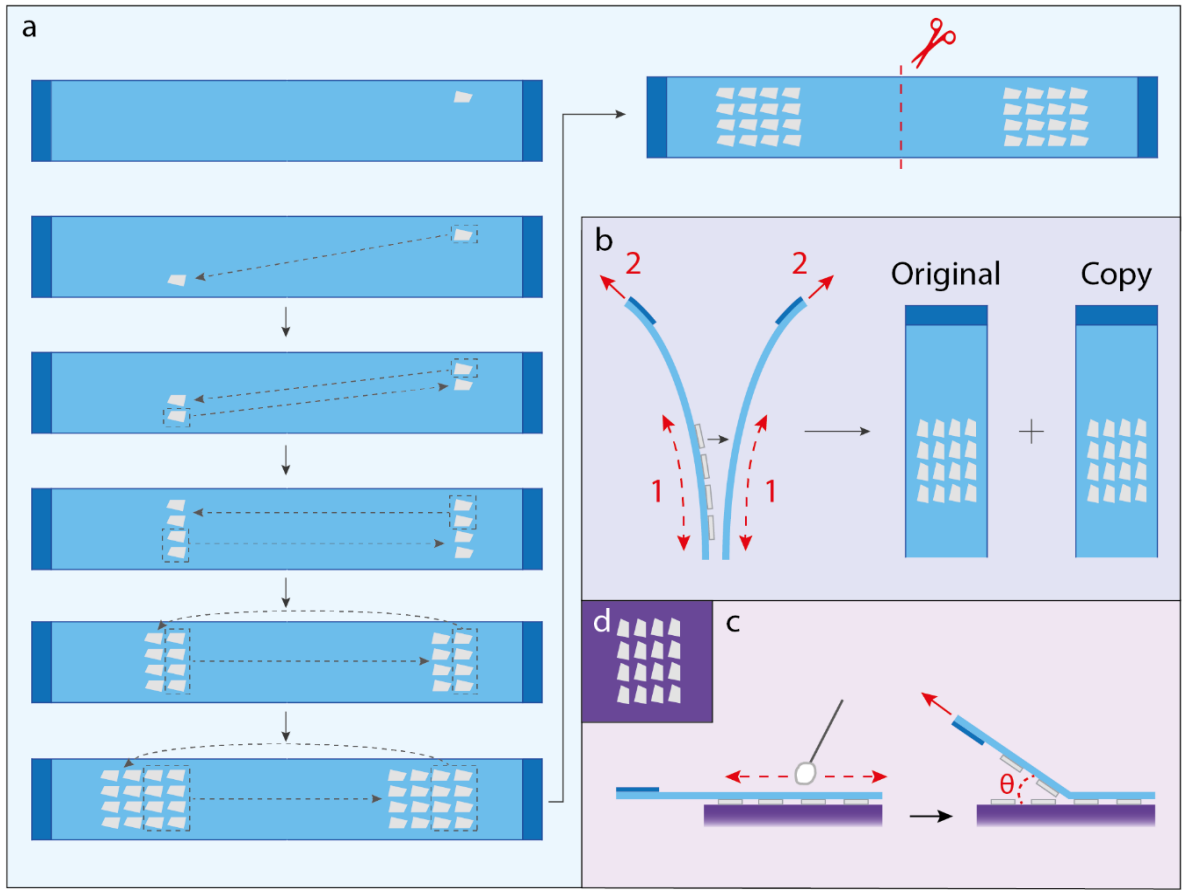

**Fig. 3 | Exfoliation workflow.** (**a**) Stamp production: fold a piece of blue tape onto itself to replicate an initial hBN flake into arrays. (**b**) Fabrication of additional stamps: rub a stamp against a new piece of tape (1) and pull apart swiftly (2). (**c**) Flake transfer to the Si/SiO$_2$ chip: use a cotton swab for gentle transfer. (**d**) Sketch of the final chip with exfoliated flakes (top view). Adapted from reference 20.

**Software**

- NanoFrazor software (NFTop 3.4.0)

- Python (v3.13)

- Pip: https://pypi.org/project/pip/

**Table 1 | Libraries used**

| Name | Module | Version |
|---|---|---|
| os | built-in | – |
| matplotlib | matplotlib | 3.9.2 |
| pickle | built-in | 4.0 |
| cv2 | opencv-python | 4.10.0.84 |
| numpy | numpy | 2.1.1 |
| math | built-in | – |
| re | built-in | 2.2.1 |
| PyQt6 | PyQt6 | 6.7.1 |
| sys | built-in | – |
| scipy | scipy | 1.14.1 |
| copy | built-in | – |
| tkinter | built-in | 0.1.0 |
| ctypes | built-in | 1.1.0 |
| sklearn | scikit-learn | 1.5.2 |
| pyinstaller | pyinstaller | 6.10.0 |

- Github Repository: https://github.com/Snunder/FunFit

## Procedure

**Grayscale nanopatterning of PPA using tSPL. TIMING: 1–3 h**

Steps 1–5 are illustrated in Fig. 2.

1) *O₂ plasma cleaning.* Clean the substrate surface (Si/SiO$_2$ chip with exfoliated hBN flakes) with O$_2$ plasma in a RIE chamber. This promotes wetting of the surface and favors uniform spin coating. (Suggested cleaning recipe: Flow rate = 10 sccm, RF power = 15 W (resulting substrate bias is 158 V), cleaning time = 30 s).
   ▲ **CRITICAL** Avoid this step if the sample is sensitive to O$_2$ plasma, *e.g.* unprotected graphene will be etched in an oxygen plasma.

2) *PPA spin coating.* Place the chip in a spinner and use an Eppendorf pipette to cover the whole chip with the PPA solution. For example, use ~50 µL for a chip of size 1.5 cm x 1.5 cm. To get a film thickness of ~80 nm for a 5 wt. % solution, set the spin program to a spin speed of 2000 r.p.m., a spin acceleration of 2000 r.p.m./s, and a spin time of 40 s. Afterwards, bake the sample on a hotplate set at 110 °C for 2 min to remove residual solvent. The thickness of a spin coated film can be estimated using established models (https://www.ossila.com/pages/spin-coating) and the PPA solutions can be prepared with different concentrations to produce different ranges of film thickness, depending on spin speed. A thicker film generally improves the film uniformity, however, it requires deeper tSPL patterning and longer etch times, both factors which may degrade the pattern quality.

▲ **CAUTION** Ensure that the hot plate is not close to waste, solvents, or other combustible materials in the fume hood.

◆ **TROUBLESHOOTING** See Table 4.

3) *Measure PPA film thickness and profile.*

<u>Measure with tSPL:</u> This measurement method allows for quick and local inspection of film thickness using tSPL. The results from this method can be corroborated by measuring a scratch in a spin coated film with AFM, which is more time consuming but also more reliable and accurate. Load the sample into the tSPL tool and write a test pattern at 1200 °C, 9 V, to push the tip to the bottom of the resist. The thickness can be estimated under such conditions as: Thickness = measured max depth + 2 nm, where the ~2 nm is leftover resist that does not decompose and sublimate. The surface topography can also be measured by the tSPL cantilever prior to patterning, allowing the user to avoid non-uniform regions during the patterning process, such as cracks and bubbles. It should be noted that films spin-coated over nanoscale objects, such as exfoliated hBN flakes, often have a non-flat topography[22], especially in the vicinity of the object edges. Therefore, patterning near edges should generally be avoided, if possible. See Extended Data section 1 and Fig. ED1.1 for details on loading the tSPL cantilever.

<u>Measure with AFM:</u> The film thickness can also be checked by making a scratch with metal tweezers in a spin-coated film and measuring the step height with AFM or profilometry. We find that these independent estimates (tSPL and AFM) generally agree within a ± 10 nm uncertainty, where the AFM measurement can be taken as the more accurate of the two methods for flat films without buried objects. For local measurements on buried objects, tSPL may provide a more accurate thickness.

◆ **TROUBLESHOOTING** See Table 4.

4) *Pattern setup and calibration*. Import the bitmap of the design pattern and assign a value to the pixel size, where 20 nm is a typical starting value. Set minimum and maximum pattern depth in the polymer. Perform a few test runs of the target pattern, allowing the read-write (Kalman) feedback loop to calibrate the writing parameters. See Box 1 for information on bitmap generation based on mathematical design functions.

◆ **TROUBLESHOOTING** See Table 4. Moreover, see Extended Data sections 1 and Figs. ED1.2–1.3 for tSPL configuration and pattern setup details. Patterning details are found in Extended Data section 2 and Fig. ED2.1.

5) *Write pattern*. Once the target and measured profiles match, abort the test run and initiate the actual pattern. Waiting longer may deteriorate the patterning conditions as more material agglomerates on the tip, altering the effective tip shape.

◆ **TROUBLESHOOTING** See Table 4 and Extended Data section 2 and Fig. ED2.1

**Pattern transfer to hBN with RIE. TIMING ~30 min**

Steps 6–8 are illustrated in Fig. 2.

6) *Pattern transfer to hBN.* Use RIE to transfer the pattern from PPA into hBN flakes. Suggested recipe: Gas type = $SF_6$, flow rate = 15 sccm, substrate bias = 40 V (results in ~24 W etch power). We found the vertical etch rate in both PPA and hBN to be approximately 1.35 nm/s, which may vary depending on the specific recipe, equipment, and experimental conditions. The etching time is usually on the order of 1–2 minutes, and the quoted step timing above includes time for pumping and venting the etching chamber, as well as sample loading and removal.

◆ **TROUBLESHOOTING** See Table 4 and Extended Data section 3 for information on how to calculate the etch time (See Fig. ED3.1) and determine the etch rate (See Fig. ED3.2).

7) *Sublimate remaining PPA.* Bake the sample on a hotplate set at 250 °C for 30 s to sublimate residual PPA.

▲ **CAUTION** Ensure that the hot plate is not close to waste, solvents, or other combustible materials in the fume hood.

8) *$O_2$ plasma cleaning.* Repeat the process in step 1. hBN should remain unaffected by oxygen plasma, and hence, the finalizing plasma clean should not affect the topography of the produced pattern. However, it is recommended that the user checks the flake thickness with AFM before and after this step to confirm that there is no residual etching.

▲ **CRITICAL** Avoid this step if the sample is sensitive to $O_2$ plasma. For example, unprotected graphene will be etched by oxygen plasma.

**Topographic characterization of hBN pattern with AFM. TIMING ~1 h**

9) *Topographic characterization of transferred pattern.* After the pattern is transferred into hBN, perform tapping-mode AFM to obtain topographic data of the patterned surface.

◆ **TROUBLESHOOTING** See Table 4.

**Software package—Data preprocessing. TIMING 1–5 min**

An overview of the *FunFit* graphical user interface (GUI) and the data preprocessing are found in Figs. 4,5.

10) *Import data*. Import tSPL topography data (*.top*) or AFM data (*.spm*) to the software.

- Open the *FunFit* GUI and click on "Load files" (See Fig. 4). Select the file to be loaded (supports NanoFrazor .top (Version 1.0) and NanoScope .spm (Version: 0x09400202) file formats).
- Data is loaded to an array and automatically saved as temporary .png and .pkl files.
- Loaded data is displayed in the main GUI window.

11) *Find structured area*. Define boundaries of patterned area.

- Click on "Find structured area" (See Fig. 4) and select two opposing corners (with two clicks) of the structured area in the pop-up window. Ensure that the selected corners lie outside the patterned area.
- A rectangle is automatically displayed to indicate the pattern area.
- If satisfied with the indicated area, click "apply". This saves the selected area and the associated data as a temporary file.

12) *Step line correction (artefact removal)*. Remove scan artefacts by median alignment of scan lines.

- Click on "Step line correction" (See Fig. 4) and select a row alignment method. The user can choose between median alignment or median of differences, where the preferred choice will depend on the dataset to be analysed.
- Both alignment methods are performed only on the unpatterned region of the surface, i.e. all data points excluding the selected pattern area.
- Click on each option to see a preview before deciding, then click "apply" to choose a method.
- The image in the main GUI window and temporary files are updated.

13) *Plane levelling (artefact removal)*. Compensate tilted scan planes by plane fitting and subtraction to recover the flat sample surface.

- Click on "Plane levelling" (See Fig.4).
- A plane is fitted to the data excluding the selected pattern area.

- The fitted plane is automatically subtracted from the data to remove tilt and set the surface zero.
- The temporary file is automatically updated.
- Plot plane fit with a sample of data points (See Fig. 5).
- Plots can be saved as .svg files by the user.

14) *Plot line-cuts.* Plot line-cuts for user visualization.

- Click on "Plot line cuts" (See Fig. 4). The plots in the points below are generated automatically.
- Plot 20 equally spaced lines of the selected pattern area after corrections.
- Plot average of all lines after corrections with standard deviation.
- Plot selected pattern area after corrections.

**Software package—Function fitting. TIMING 1–30 minutes**

See Figs. 5–9 for an illustrative overview of the function fitting and analysis capability in *FunFit*.

15) *Function fitting.* Fit preset functions or custom user-defined functions to surface topography data using nonlinear least squares fitting.

- Repeat step 11, but this time ensure that the selected corners are inside the patterned area.

- Click on "Fit functions" (See Fig. 4) and choose one of two fitting options (i) fit a sinusoidal superposition in a rotational basis "quasicrystal" (N>=1 cosines arranged with equal angle spacing in plane), or (ii) fit a custom function.
- For a quasicrystal, the function to be fitted is shown. Input an initial guess for all variables, where the target parameters for the bitmap function are a good starting point. Note that the amplitude and the offset may vary significantly from the target, while the period is usually similar.
- For a custom function, input an arbitrary function when prompted to do so. Then additional input boxes appear, asking for initial guesses for all variables.
- If the selected function is non-periodic, a window containing the data appears. Here, select the origin (center) of the function.
- If the initial guesses are sufficiently close and the roughness is sufficiently small then a fitted model is created. The tolerance on initial guesses and roughness and success of the fit is typically different for different datasets.
- The raw data, fitted model, residuals, and FFT plots are shown in a single figure with the RMSE of the residuals displayed as a metric to quantify topographic fidelity (See Figs. 6–9 and Tables 2,3). The fitted parameters are displayed as numbers above the plots.
- The figure is saved as a .svg file by the user.
- Tables 2,3 show the fit parameters for the sinusoidal and quasicrystal functions measured with tSPL in PPA before pattern transfer, and measured with AFM in hBN after pattern transfer. The RMSE error is evaluated, and these numbers are compared with the target function.

16) *Roughness analysis*

- Click on "Roughness of flat area" (See Fig. 4a) and select opposing corners of area to be used for roughness analysis.
- A plane is fitted to the selected area and a plot of the residuals is shown with RMSE value.
- The figure is saved as a .svg file by the user.

---

Box 1: Bitmap generation

- *FunFit* also provides the option to generate a bitmap from a mathematical design function, if one is not already available. This bitmap can then be imported into the tSPL system in step 4.
- This capability is enabled by clicking on the "Bitmap generator" in the *FunFit* GUI (See Fig. 4), choosing the function to be created, and inputting values for the pixel size and overall pattern size. Note that the pixel size entered here should match the pixel size set in the NanoFrazor tool during patterning.
- The user is prompted for a folder location to save the bitmap.

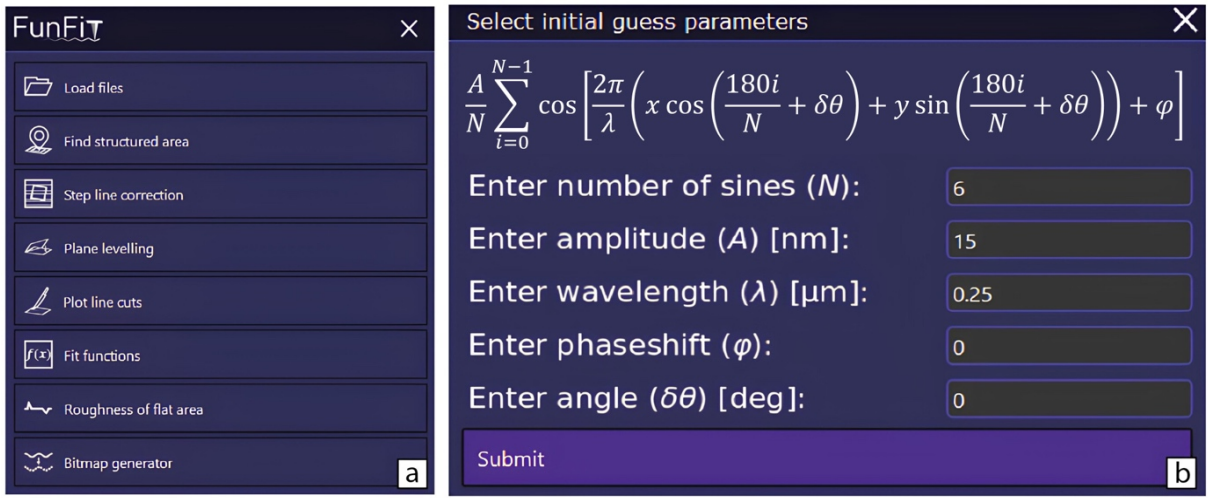

**Fig. 4 | Overview of *FunFit* GUI. (a)** Graphical user interface (GUI) for the *FunFit* software package. **(b)** Function fitting window where the user is prompted to provide initial guesses for the fit parameters of their function.

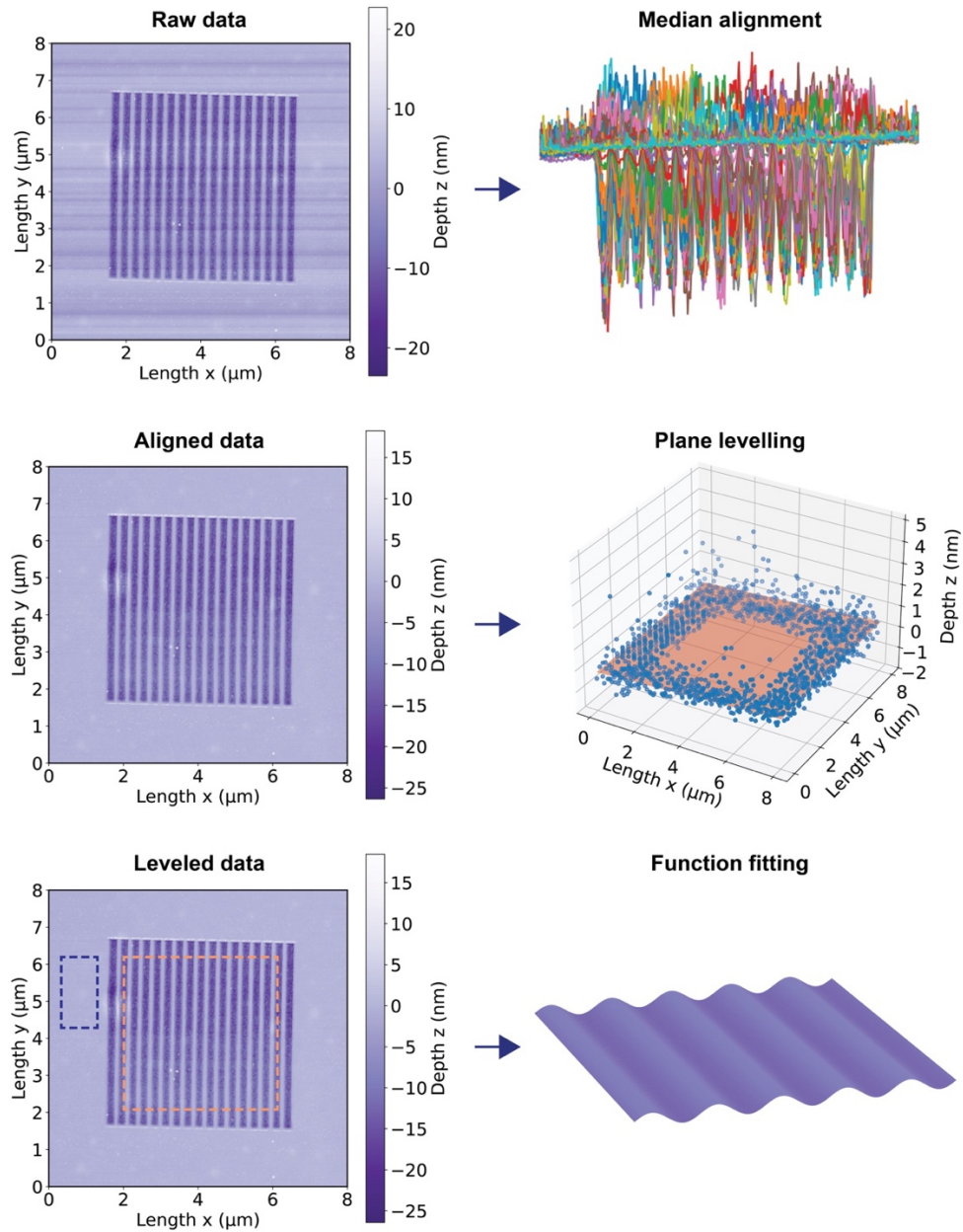

**Fig. 5 | Overview of *FunFit* features.** Raw data is treated with a median alignment algorithm to generate aligned data, where the median alignment panel shows multiple individual line cuts overlayed after median alignment. The aligned data is treated with plane fitting and subtraction to level the data to a flat surface. The levelled data is then used for function fitting and analysis. The dashed orange box marks the area used for function fitting, the dashed blue box marks the area for surface roughness analysis. Both areas are user defined.

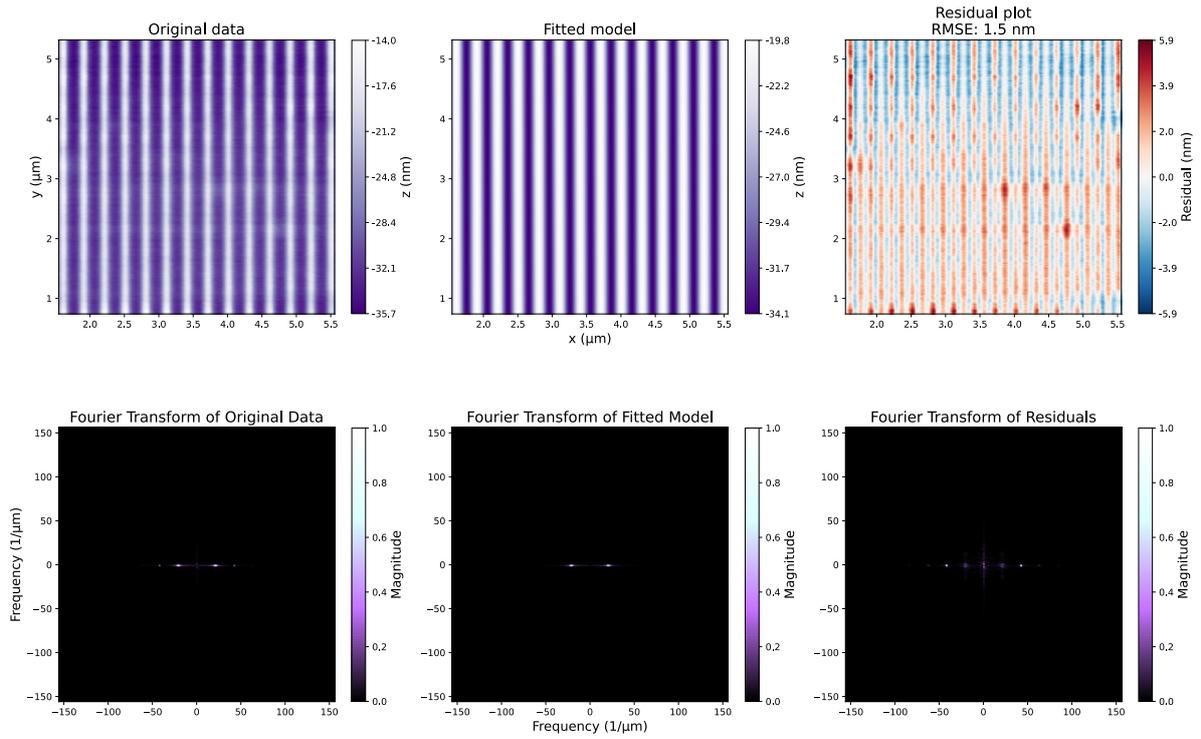

**Fig. 6 | Analysis of sine pattern in PPA measured by tSPL. Top Row**: Left: Surface topography of PPA measured by tSPL, Middle: fitted model, Right: residuals plot to quantify error as RMSE. **Bottom Row**: Fast Fourier transform (FFT) of the data presented in the panels above.

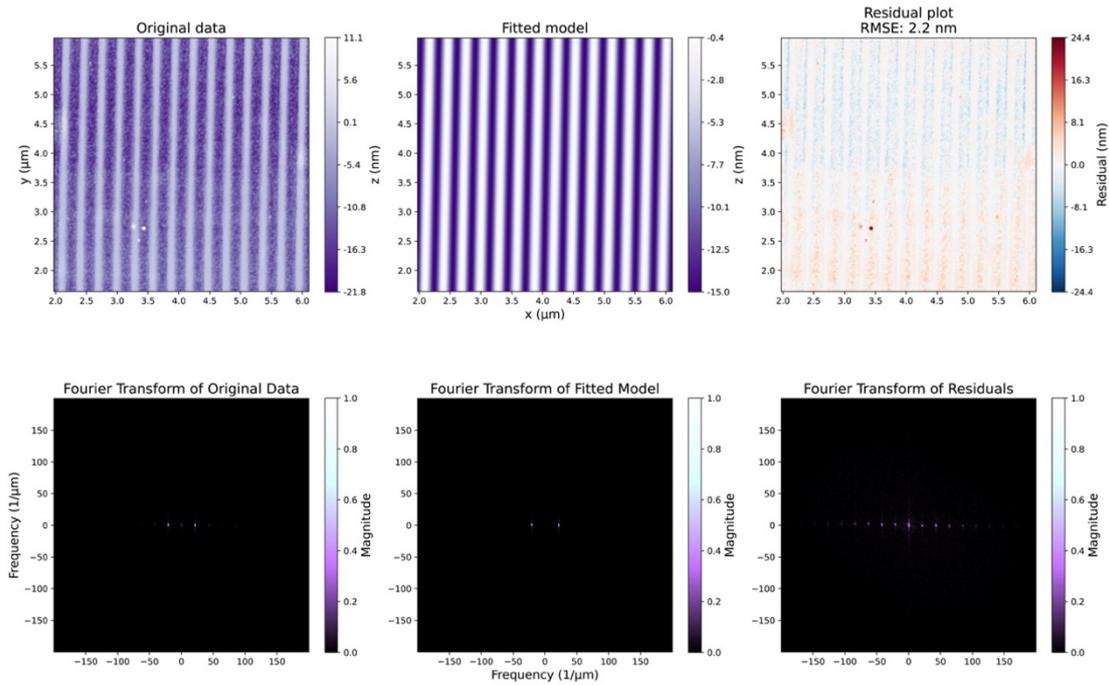

**Fig. 7 | Analysis of etched sine pattern in hBN measured by AFM. Top Row**: Left: Surface topography of hBN measured by AFM, Middle: fitted model, Right: residuals plot to quantify error as RMSE. **Bottom Row**: Fast Fourier transform (FFT) of the data presented in the panels above.

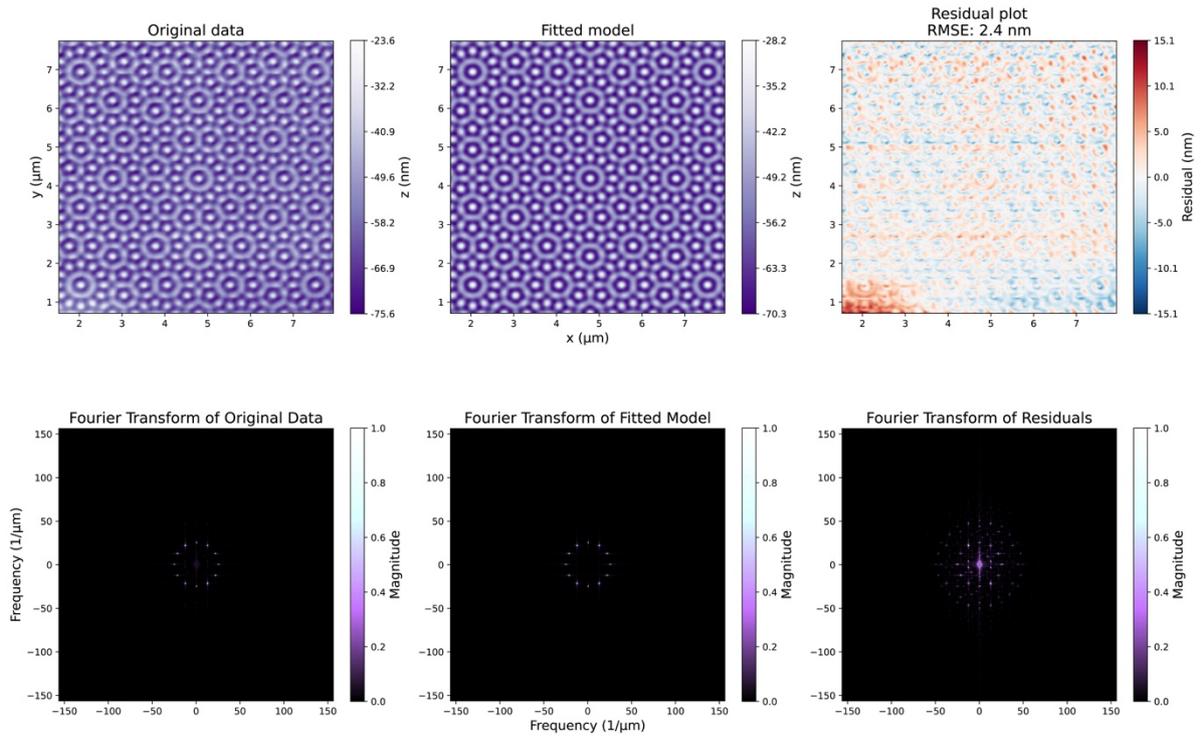

**Fig. 8 | Analysis of quasicrystal pattern in PPA measured by tSPL. Top Row**: Left: Surface topography of PPA measured by tSPL, Middle: fitted model, Right: residuals plot to quantify error as RMSE. **Bottom Row**: Fast Fourier transform (FFT) of the data presented in the panels above.

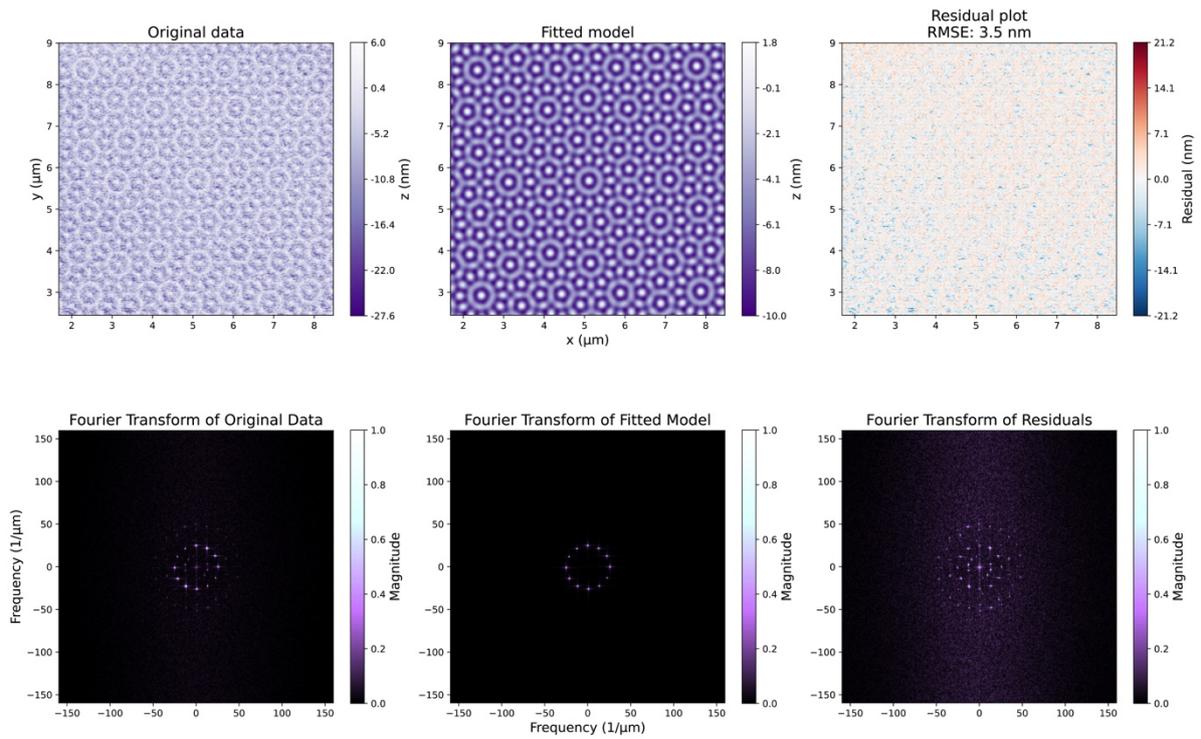

**Fig. 9 | Analysis of etched quasicrystal pattern in hBN measured by AFM. Top Row**: Left: Surface topography of hBN measured by AFM, Middle: fitted model, Right: residuals plot to quantify error as RMSE. **Bottom Row**: Fast Fourier transform (FFT) of the data presented in the panels above.

**Table 2 | Fit parameters for sinusoidal function (Fig. 6,7).**

| Sine | bitmap | PPA | hBN |
|---|---|---|---|
| $A$ (nm) | 7.5 | 6.9 | 7.3 |
| $\lambda$ (nm) | 300 | 300 | 296 |
| $\varphi$ (deg.) | 0 | 1.2 | −2.5 |
| $\delta\theta$ (deg.) | 0 | −0.0077 | 1.0 |
| $\Delta$ (nm) | −27.5 | −27 | −4.7 |
| RMSE (nm) | 0 | 1.5 | 2.1 |

$$f(x,y) = A\cos\left(\frac{2\pi}{\lambda}[x\cos(\delta\theta) + y\sin(\delta\theta)] + \varphi\right) + \Delta$$

**Table 3 | Fit parameters for quasicrystal function (Fig. 8,9).**

| Quasicrystal ($N$ = 6) | bitmap | PPA | hBN |
|---|---|---|---|
| $A$ (nm) | 20 | 28 | 7.7 |
| $\lambda$ (nm) | 250 | 250 | 247 |
| $\varphi$ (deg.) | 0 | −0.043 | 0.20 |
| $\delta\theta$ (deg.) | 0 | −0.0074 | −1.0 |
| $\Delta$ (nm) | −50 | −56 | −5.8 |
| RMSE (nm) | 0 | 2.5 | 3.5 |

$$f(x,y) = \frac{A}{N} \sum_{i=0}^{N-1} \cos\left(\frac{2\pi}{\lambda}\left[x \cos\left(\frac{180i}{N} + \delta\theta\right) + y \sin\left(\frac{180i}{N} + \delta\theta\right)\right] + \varphi\right) + \Delta$$

◆ **TROUBLESHOOTING** See Table 4 and Extended Data.

**Table 4 |** Troubleshooting.

| Step | Problem | Possible reason | Solution |
|---|---|---|---|
| 2 | Resist does not wet the surface, resulting in a non-uniform film (or no film at all). | Sample surface is not completely clean and may have a contamination layer or unfavourable surface chemistry for wetting with PPA in anisole. | Repeat oxygen plasma clean. If that doesn't help, consider making a new solution of PPA. |
| 3 | Estimated film thickness is different from expected thickness. | The PPA thickness may vary across the substrate, especially in the region where exfoliated 2D flakes reside. | Check PPA thickness in a few places, most importantly locally in the area of interest, e.g. on top of the exfoliated 2D flake. Cross-check with AFM for a spin-coated film on a flat substrate as a benchmark. |
| 4–5 | Measured pattern profile is too shallow (deep) compared to the target profile. | Temperature or write force is too low (high). | Increase (decrease) temperature or write force. The temperature typically ranges between 950 °C and 1200 °C. The max force is 9.5 V, but typical write forces are ~3–7 V. |
| 4–5 | The measured pattern profile does not match the target, even though the feedback and depth ranges have converged. | This may be caused by a tip with a non-ideal shape due to contamination accumulated during the patterning process. | Write a few more test patterns to see if it improves, and if not, consider changing the tip. |
| 4–5 | Flat pattern valleys in the measured profile compared to the target profile. | Too shallow of a vertical distance between the pattern and the substrate alters thermal flux from the tip to the substrate, hindering patterning. | Leave 10-50 nm vertical distance between the target pattern valleys and the substrate, a "cushion" (See Fig. ED3.1). |
| 6–9 | The etch rate(s) appear different than stated in the procedure. | Chamber conditions may be different. | Perform preliminary tests to estimate the precise etch rate in PPA and/or hBN in the RIE chamber (See Fig. ED3.2). |
| 9 | Pattern is not transferred fully into the dielectric. | Underetch. | Evaluate etch rate and increase etch time accordingly. |
| 9 | Color change of the flake area surrounding the pattern + the unpatterned regions are rough in the AFM scan. (Fig. 2 bottom-right panel). | Overetch. | Evaluate etch rate and decrease etch time accordingly and/or in the pattern setup in the tSPL system, increase the vertical offset of the minimum pattern depth into the PPA layer (etching buffer) to minimize etching of the surrounding area in the case of a small over etch (See Fig. ED3.1). |

**TIMING**

- Steps 1–5, Grayscale nanopatterning of PPA using tSPL: 1–3 h
- Steps 6–8, Pattern transfer to hBN with RIE: ~30 min
- Step 9, Topographic characterization of hBN pattern with AFM: ~1 h
- Steps: 10–14, Software package–Data preprocessing: 1–5 min
- Steps: 15–16, Software package–Function fitting: 1–30 mins

**Anticipated results**

The protocol is composed of simple steps with clear visual clues for inspection at every stage in the procedure (See Fig. 2). It is therefore anticipated that an inexperienced user who has become familiar with all steps of the protocol should be able to perform it within a single working day, producing for instance an hBN flake with a smooth topographic pattern etched into the surface. However, substantial process development and optimization may be required to implement this protocol for applications that require fabricating surfaces with optimized profiles and low roughness.

Once the hBN flake is patterned, the surface topography is measured using AFM, providing a dataset to compare to the target mathematical function and the intermediate tSPL data. The data is analysed in the *FunFit* software package, which generates fits that provide quantitative characterization of the pattern topography, the error, and the roughness or any other artefacts introduced by the fabrication procedure (See Figs. 4–9, and Tables 2,3). The quantitative information produced by the analysis is valuable to assess the fidelity of patterned topographic landscapes, which can be used as an input to theoretical models that calculate the influence of the landscape on a particular experimental system. Furthermore, the information provided by the code can be used to assess the nanofabrication and pattern transfer process, offering a feedback mechanism by which to reduce errors, minimize roughness, and optimize the results of the procedure. Once a process is optimized, this protocol offers a standardized and reliable framework by which to assess reproducibility[23] over longer time scales and between varying laboratory environments.

**Author contributions**

N. L. conceived the project and designed the protocol. M. N. and S. L. wrote the code with input from N. L., C. H. S., and R. J. C. H. S., D. H. N., and M. N. exfoliated hBN. N. L., C. I., and C. H. S. patterned polymer resist with tSPL. N. L., C. H. S., and M. N. etched hBN. C. H. S., M. N., and N. L. characterized hBN with atomic force microscopy. C. H. S., M. N., and S. L. analysed the data with input from N. L. C. H. S., M. N., S. L., and N. L. produced the figures. C. H. S. and N. L. wrote the manuscript with input from all authors. N. L., S. R., P. B., and T. J. B. supervised the project.

**Acknowledgements**

N. L. acknowledges funding from the Swiss National Science Foundation (*Postdoc Mobility* P500PT_211105) and the Villum Foundation (*Villum Experiment* 50355). T. J. B. and P. B. acknowledge support from the Novo Nordisk Foundation (BIOMAG NNF21OC0066526).

**Competing interests**

The authors declare that they have no competing financial interests

**Correspondence and requests for materials** should be sent to Nolan Lassaline at nlasso@dtu.dk.

Asselberghs, I., Barkan, T., Taboryski, R. & Pollard, A. J. Closing the reproducibility gap: 2D materials research. arXiv:2409.18994 (2024).

# Extended data

**Extended data section 1: tSPL—Setup and calibration**

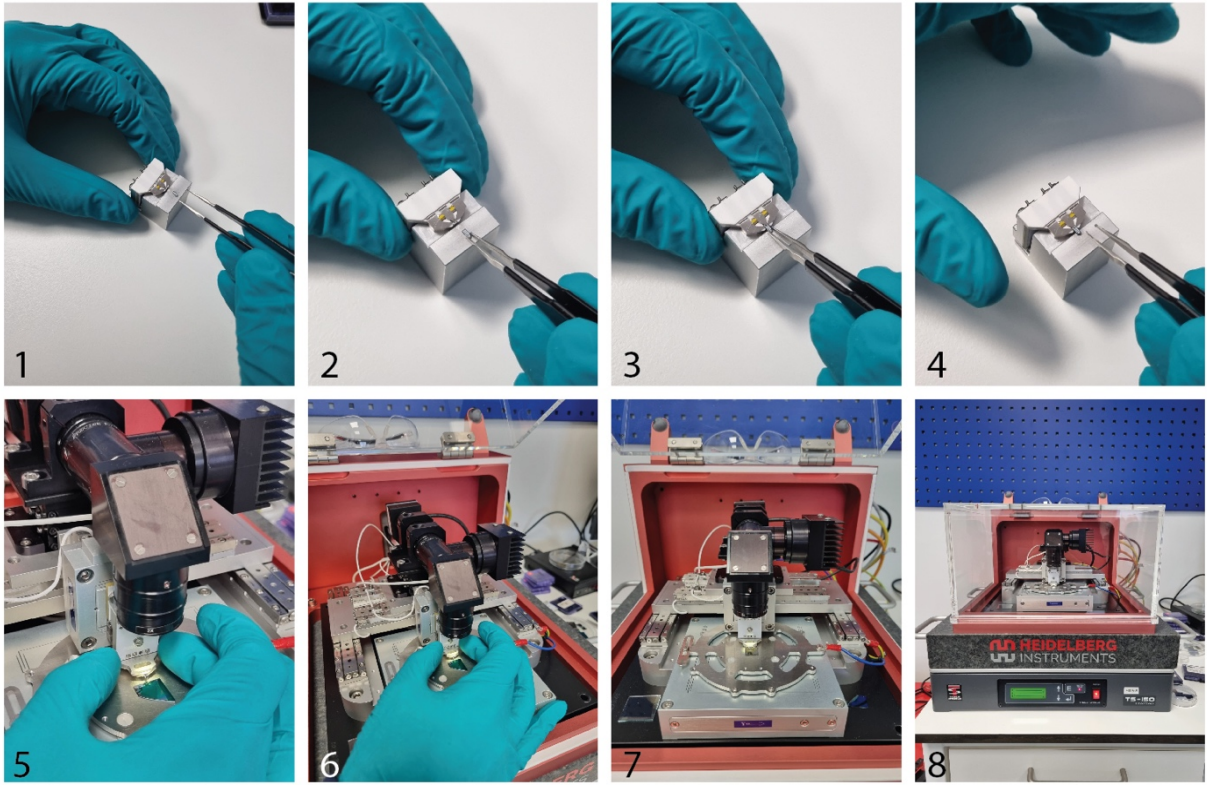

**Fig ED1.1 | Loading a cantilever for tSPL. 1**–**4**, Loading the cantilever into the cantilever holder by pressing the holder down with one hand and using tweezers to insert the cantilever into the holder with the other hand. Both hands let go when complete. **5**–**8**, Attaching the cantilever holder to the NanoFrazor scan head and closing the housing. The user approaches the tip to the sample below, where the tool is ready for configuration.

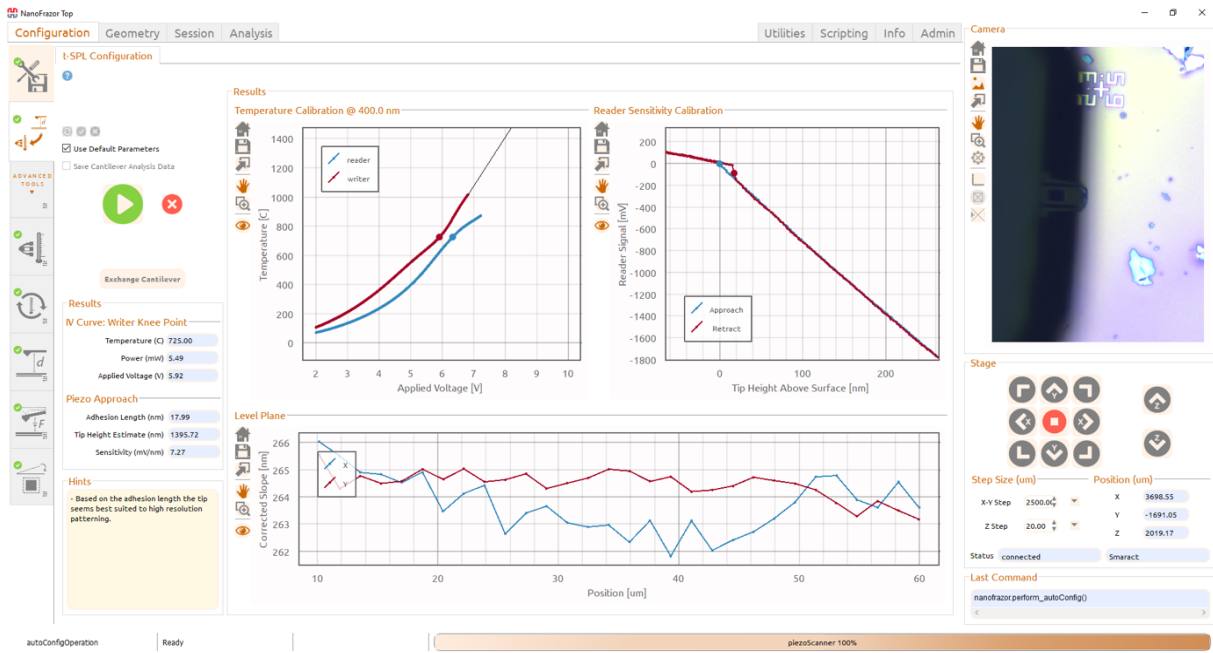

**Fig ED1.2 | Configuration.** Screen capture of the NanoFrazor software displaying typical results of the configuration procedure. The configuration is simply run by pressing the green play button once the tip has been brought close (<100 μm) to the sample surface.

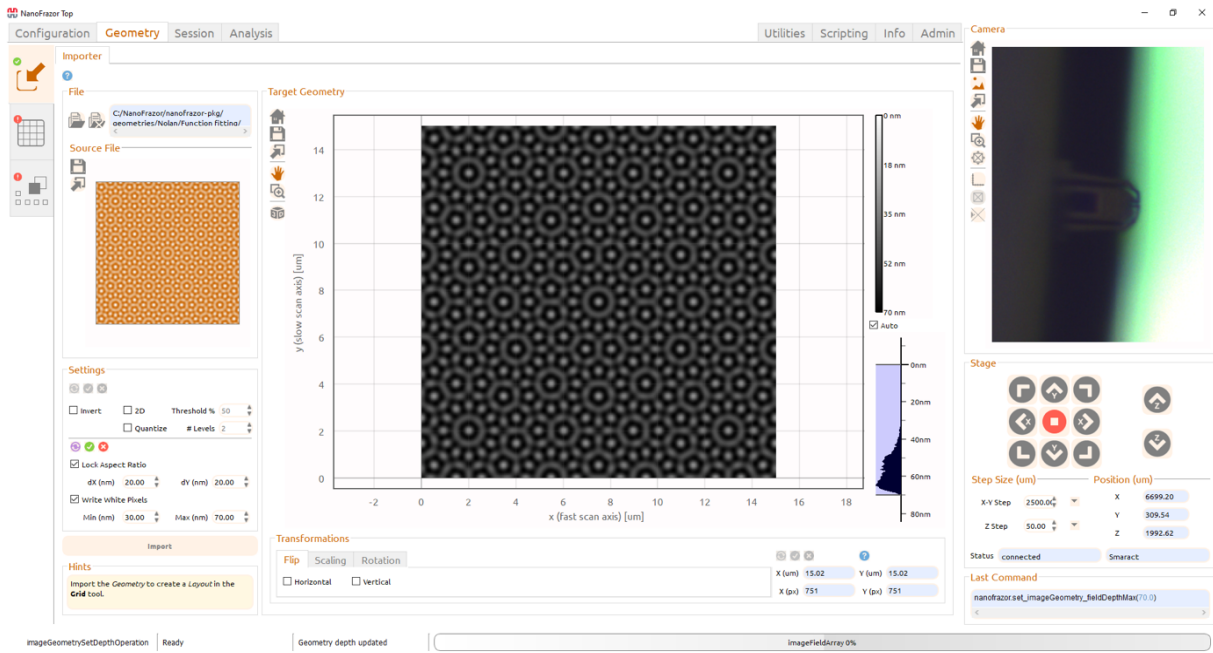

**Fig ED1.3 | Import pattern geometry and set dimensions.** Screen capture of the NanoFrazor software displaying a typical setup of pattern geometry. Here, a bitmap is loaded and the pixel size is set in the NanoFrazor software, where 20 nm × 20 nm is a typical value. The user then sets the depth range for the pattern. Here, the range is set to 30–70 nm into the surface to provide 40 nm of modulation depth and 30 nm of buffer for etching. The pattern is then imported.

## Extended data section 2: tSPL—Patterning

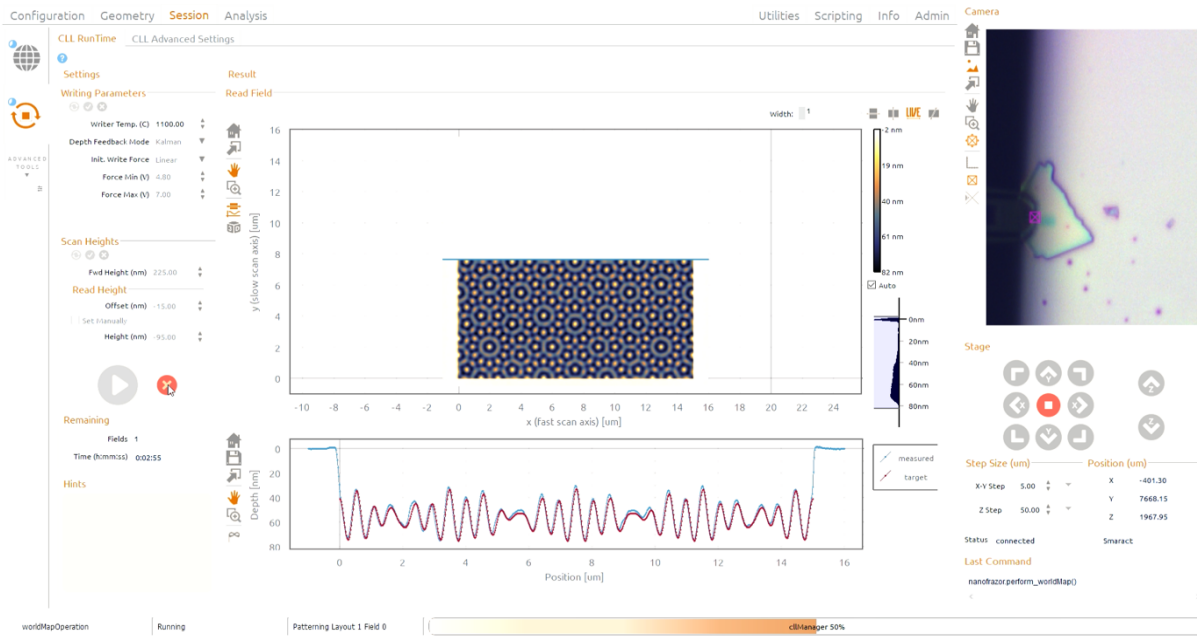

**Fig ED2.1 | tSPL patterning.** Screen capture of the NanoFrazor software displaying the patterning process of a quasicrystal structure into a PPA film. The top right image is an optical microscope looking down on the cantilever from the top, here showing the cantilever patterning a PPA film spin-coated on top of a hBN flake. The pattern is fabricated and measured simultaneously, where the tSPL probe acts as a combined read–write lithography and imaging tool. The measured topography is shown in the middle, and a line-by-line comparison between the target surface and measured profile is shown below. The pattern here is performed with a set temperature of $T$ = 1100 °C, a user-defined force initialization range of 4.8–7.0 V, a cantilever forward height of 225 nm, and Kalman Feedback and Bowing Correction are turned on.

**Extended data section 3: Etching**

The etch time can be calculated by the following formula:

$$\text{Etch time} = \text{strike time} + \frac{\text{cushion} + 2A + \Delta}{\text{rate}_{\text{PPA}}}$$

Where the strike time is how long it takes for the plasma to be generated after the etching recipe is started (~7 seconds in our system), the cushion and pattern amplitude *A* is indicated in Fig. ED3.1 (typically in units of nm), $\Delta$ is a user-defined buffer (typically in nm) that can be positive or negative to engineer a "miss" to one side of the under-etch–over-etch balance if it is critical for the process, and $\text{rate}_{\text{PPA}}$ is the etch rate of PPA (typically in units of nm/s).

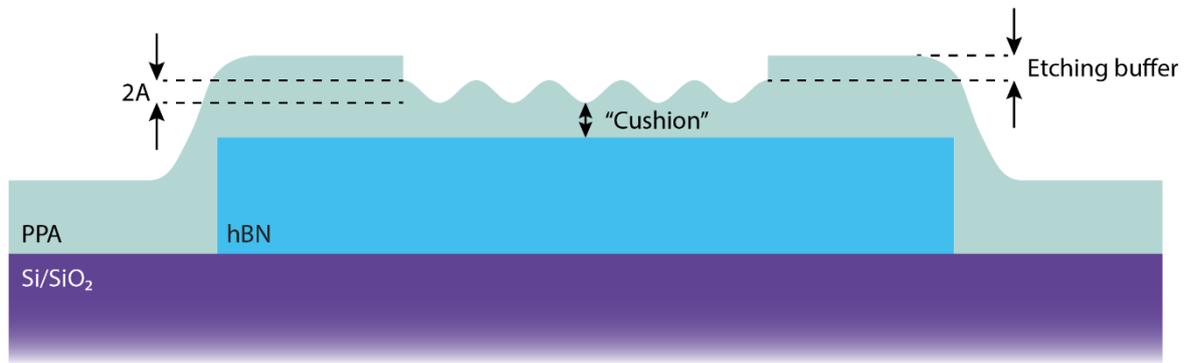

**Fig ED3.1 | Geometry to be etched.** Illustration of the cross-section for a patterned PPA layer spin-coated on a hBN flake.

The etch rate of PPA and the underlying material is a critical parameter that is sensitive to several environmental factors. The etch rate can be determined by etching a spin coated film for some time and measuring the thickness of the film before and after etching. The difference in thickness for the etch time can be used to calculate the etch rate. The rate varies in different etching systems in different laboratories, so a useful strategy to determine an accurate etch rate is to start by assuming a PPA etch rate of ~1 nm/s. The thickness of a spin coated film can then be measured, and the etch time can be calculated to etch away 50% of the film, based on the assumed etch rate. After etching, the thickness of the film can be measured by AFM and the new etch rate can be determined to update the etch rate (See Fig. ED3.2). This process can be repeated until it converges to a more precise number (1.35 nm/s in our system). The *FunFit* software package can be used to determine etch rates and pattern amplification/compression factors for when the patterns in PPA are transferred to other materials such as hBN.

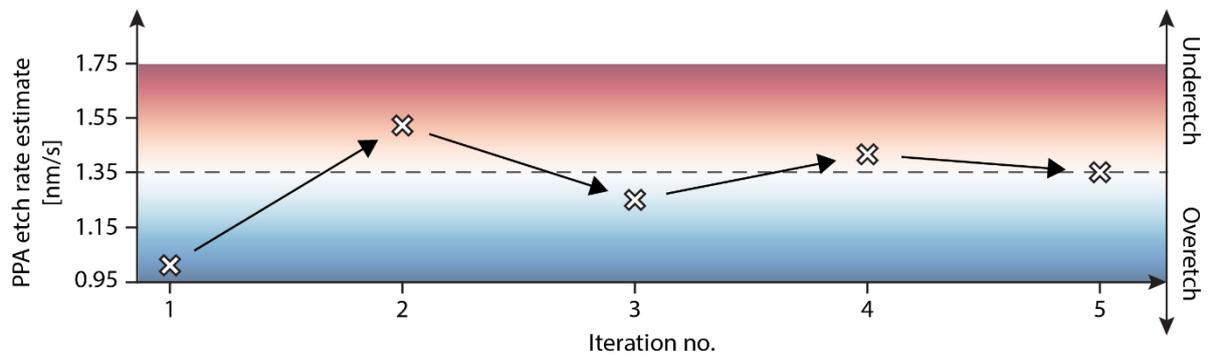

**Fig ED3.2 | Determining the etch rate.** Schematic showing the approach to determine the unknown etch rate of a particular system. The etch rate is guessed to be ~1 nm/s, and by measuring the film thickness before and after etching with AFM, this guess can be applied to etch a film approximately in half. The etch rate can be calculated from this experiment, providing an updated guess for the etch rate. Due to natural deviations in determining the etch rate, this process can be repeated until it converges on a representative etch rate of a particular system in a particular state. For our system the etch rate of PPA was determined to be 1.35 nm/s using 15 sccm of $SF_6$ gas and a substrate bias of 40 V.

# Source code and data availability

Data and code are freely available at GitHub: https://github.com/Snunder/FunFit